# Small Bodies Tell the Story of the Solar System: A Scientific Rationale for a Multi-Target Small Body Sample Return Program including the Earth-based Laboratory Analysis of Returned Samples

Primary Authors:
Seth Jacobson, Michigan State University, 1-517-355-1941, seth@msu.edu
Maitrayee Bose, Arizona State University, 1-480-965-4244, Maitrayee.Bose@asu.edu

Co-Authors:
Dennis Bodewits, Auburn University
Marc Fries, NASA Johnson Space Center
Devanshu Jha, MVJ College of Engineering
Prajkta Mane, Lunar and Planetary Institute
Larry Nittler, Carnegie Institution of Washington
Scott Sandford, NASA Ames Research Center
Michelle Thompson, Purdue University

**Compelling Reasons For Small Body Exploration**

Small bodies are time-capsules of different eras of solar system history from the most primitive materials within the solar system to evolved pieces of larger bodies. While small bodies include asteroids, comets, dwarf planets, Trojans, Centaurs, trans-Neptunian objects, small satellites, and interplanetary dust, a diverse selection of small bodies either reside near Earth or have been scattered by the gravity of the planets onto near-Earth orbits, ultimately making all parts of the solar system accessible to exploration. Furthermore, the surfaces of many small bodies can be affordably accessed with spacecraft due to their low gravity and non-existent or sparse atmospheres. Lastly, while our meteorite collections are an important source of knowledge about solar system history, they are a biased and incomplete representation of small body astromaterials.

Small bodies record the radial compositional gradients of material that were present in the protosolar disk, and they represent all stages of the formation and early evolution of the solar system. Primitive small bodies are the debris left over from planet formation and contain examples of the primordial ingredients from which the planets and life arose. Small bodies record internal processing such as aqueous alteration, thermal metamorphism, melting and differentiation. These evolved small bodies record processes that occurred during the formation and evolution of planets. Thus, small bodies trace growth from primordial condensates and presolar/interstellar grains to pebbles, planetesimals, and planetary embryos to planets. Since catastrophic collisions can completely shatter protoplanets, small bodies can be samples of core, mantle, or crustal material of once much larger bodies. Indeed, this is the only known reservoir of accessible core and, potentially, deep mantle material. Small bodies also record the history of the solar system such as the dynamical evolution of the solar system, the evolution of surface materials through time and as they approach the Sun, and the primordial cosmochemical gradients established within the solar nebula. Samples from small bodies in the solar system represent a diverse set of materials, and study of these materials in terrestrial laboratories can provide information about the history and evolution of the solar system in ways that cannot be determined by remote and in situ observations.

Small bodies from throughout the solar system have been scattered onto orbits that are accessible by spacecraft from Earth on reasonable mission timescales and budgets. Near-Earth asteroids are sourced from throughout the Main Belt, although parts of the Main Belt itself can be directly accessible for sample return using launch vehicles in development. The moons and Trojans of Mars are also directly within reach. Jupiter-family and long-period comets can have perihelions within 1 AU and are representative examples of the Kuiper Belt and Oort Cloud, respectively. NASA, ESA and JAXA have a long history of successful operations around small bodies. Sample return missions to small bodies have historically been more limited, but past and current missions include NASA's *Stardust* and *OSIRIS-REx* missions and JAXA's *Hayabusa* and *Hayabusa2* missions.

Because of their low gravity and lack of atmospheres, many small bodies are accessible using spacecraft missions in the Discovery and New Frontiers classes, as demonstrated by completed missions such as *Near-Earth Asteroid Rendezvous*, *Stardust*, *Deep Impact,* and *Dawn,* current missions such as *New Horizons* and *OSIRIS-REx,* as well as future missions such as *Psyche* and *Lucy*. These missions have shown that exploration of small bodies can provide an extremely effective means of learning the story of the origin and evolution of the solar system and demonstrate the potential value provided by future missions. In particular, the science returned by these missions have opened new questions, some of which can only be approached with the in-depth investigation of returned samples.

**Sample Return Missions Are The Apex Of Small Body Exploration**

The science goals of small body exploration are to establish the early conditions of the solar nebula, understand how materials evolved during the stages of planetary growth, and determine the source of the building blocks of Earth including water and other ingredients necessary for life. These goals are responses to the fundamental questions: 'What is out there in the universe?', 'What is the history of Earth and the solar system?', and 'How did life arise on Earth and may it have arisen elsewhere?' that not only drive the public imagination but have historically driven significant investments in science. While in-situ spacecraft exploration provides detailed information that cannot be obtained from remote telescopic observations, small body sample return missions are the only mechanism to fully connect astrophysical observations, in-situ spacecraft exploration, and meteorite analysis.

Here, 'sample return mission' refers to the acquisition of material (dust particles, surface material, drill cores, captured gases, ices, etc.) by a collecting device on a spacecraft, their return to Earth, and their analysis in a laboratory on Earth. Sample return missions and supporting infrastructure (cameras onboard the spacecraft, ground-based telescopes, etc.) must also characterize the local, regional, and global context of the sampled location.

Of all investigative methods, sample return missions provide the deepest insight into a planetary body, because returned samples can be analyzed by the most advanced laboratory techniques in terrestrial laboratories with the highest precision and detection limits and which are not achievable in remotely operated planetary missions, while still in the context of landed and remote sensing observations. This effectively allows the mission's payload to consist of all the world's analytical instrumentation. First, many of these analyses cannot be performed in space due to technological or cost limitations. Second, Earth-based laboratories can achieve measurement accuracies and precisions superior to in situ measurements, often with very small sample quantities. Third, returned samples can be and have historically been partially preserved for study in the future by methods that have yet to be developed. Earth-based laboratory analysis of returned samples from small bodies provides detailed information regarding the sample's geologic characterization, petrology, and mineralogy. Moreover, laboratories can perform high precision geochemical measurements including molecular, elemental, and isotopic compositions, which can be used for radiometric dating. Using these rich datasets, the history of both the individual parent body as well as the population they represent, can be determined.

Characterization of returned samples from small bodies can also contextualize laboratory analyses of meteorites and interplanetary dust particles (IDPs) collected on Earth. Laboratory analyses of extraterrestrial materials retrieved on Earth have, and continue to, provide answers to fundamental questions about the early solar system, but because in most cases the sources of the materials cannot be directly determined, their properties cannot be fully put into their parent body context. Returned samples overcome this issue and provide sample-asteroid connections and parent body context without ambiguity. For instance, Stardust connected IDPs to cometary dust and Hayabusa connected ordinary chondrite meteorites to S-type asteroids. Thus, by collecting samples with known provenance, sample return missions not only produce unique discoveries of their own, but also enrich the value of laboratory meteorite and dust particle studies by revealing connections between these unique sample sets.

Ground-based telescopic observations of small bodies and astrophysical disks (protoplanetary and debris), spacecraft-based remote and in-situ exploration, and theoretical modeling are enriched by sample return investigations. For instance, competing theoretical models of planet formation can now satisfactorily explain modern-day astrophysical properties such as masses and orbits of planets and small bodies, but have significantly different hypotheses regarding the histories of growth and evolution of those bodies as recorded in their chemical and isotopic compositions. Returned samples provide definitive evidence to test these hypotheses and break existing model degeneracies. Similarly, degeneracies between interpretations of remote observations either from astrophysical telescopes or in-situ spacecraft can only be broken by laboratory investigations that, due to sample preparation and analysis requirements, must occur on Earth. Astrophysical observations, in-situ spacecraft investigations, and Earth-based laboratory analyses are distinct nodes in a continuum of understanding small bodies that become tightly linked by a small body sample return program, enhancing the scientific return of all.

**Small Body Sample Return Early Success And Progress**

Despite the lack of existence of a defined small body sample return program, two successful missions have returned samples from small bodies: NASA's *Stardust* Discovery mission, which returned both dust particles from comet 81P/Wild 2 and interstellar dust, and JAXA's *Hayabusa* mission, which returned surface samples from asteroid 25143 Itokawa. These two missions clearly illustrate the power of sample return as both have provided unprecedented constraints on specific small bodies and on their relation to extant meteoritic materials. With respect to *Stardust*, the collection, return, and terrestrial analysis of a few thousand particles from the coma of Wild 2 demonstrated that comets - generally thought to have formed in the outer protosolar nebula - nonetheless contain materials that must have been formed and altered in both the inner and outer parts of the protosolar nebula. This provides clear proof that comets are not, as once thought, pristine collections of interstellar materials and that dust was mixed throughout the entire extent of the disk. Furthermore, laboratory analyses confirmed the long suspected connection of many IDPs to cometary sources. Similarly, with respect to JAXA's *Hayabusa* mission, laboratory analysis of a few nanograms of Itokawa particles, tens to hundreds of microns in diameter, confirmed a long-suspected connection between S-type asteroids and ordinary chondrite meteorites. This analysis also revealed a rich and complex history of the asteroid's parent body formation and evolution over billions of years and provided unprecedented information about space weathering processes on asteroids. The measurements that led to all of these discoveries could never have been made by remote or in situ instruments, but instead required the state-of-the-art instrumentation available in terrestrial laboratories, much of which wasn't even developed at the time the missions were designed and launched.

Two ongoing sample return missions include JAXA's *Hayabusa2* mission, which is returning with surface samples from 162173 Ryugu, and NASA's *OSIRIS-REx* mission, which is in the process of collecting surface samples at 101955 Bennu. These missions targeted C-type asteroids to test the hypothesized connection to carbonaceous chondrite meteorites. The returned samples will have known provenance, rock surfaces unaffected by passage through Earth's atmosphere, and are predicted to contain examples of the primitive material that delivered water and carbon to Earth. While there are distinct differences in hydration as observed by remote spectroscopy, the overall similarity between the two targets creates an opportunity to compare and contrast the returned samples in the context of individual parent body evolution. Overall, these missions demonstrate the desire for returned samples, and prove, beyond doubt, the technical feasibility of sample return missions from small bodies.

Beyond the selection of missions for flight, NASA has also recognized the importance of supporting sample-return science with the development and advancement of laboratory methods for eventual analysis of returned samples. NASA initiated the Sample Return and Laboratory Instrumentation and Data Analysis Program (SRLIDAP), now the Laboratory Analysis of Returned Samples (LARS) program, that funded both PI-built advanced instruments such as MegaSIMS, SARISA, and CHILI, as well as the acquisition of commercial instruments. NASA's PSD has further supported acquisition of commercial instrumentation (e.g., electron microscopes, mass spectrometers, FTIR, Raman, etc) for individual PIs and facilities through the Planetary Major Equipment (PME) program, in many cases expressly to support analysis of returned samples (as well as meteoritic and/or terrestrial samples). These instruments are available for individual PI use at a variety of institutions, including universities, research institutes, and National Labs or as part of user facilities available to the community at large. Large scale user facilities such as synchrotron X-ray and neutron beam sources are also available and have been extensively used for returned sample analysis. All this unique instrumentation is currently being used but needs continued funding for development work and updating existing instrumentation to satisfy the needs for future returned sample analyses (for an extensive review see the National Academies Report, "Strategic Investments in Instrumentation and Facilities for Extraterrestrial Sample Curation and Analysis").

**Example Science Goals Of Small Body Sample Return Missions And Earth-based Laboratory Analysis**

While the history of small body sample return and analysis demonstrates high scientific returns, the scope of existing returned samples only begins to fill the domain of possible science. As described above, small body exploration science goals encompass the entire history of the solar system, and samples returned from small bodies, likewise, cover this incredible territory. To understand the early conditions of the solar nebula, sample return mission targets may include comets and undifferentiated asteroids. To understand planetesimal growth and internal processing, targets may include larger asteroids such as Ceres or their fragments that show evidence of aqueous alteration, thermal metamorphism, and other processes at their surfaces. To understand planet formation and evolution, targets may include asteroids that appear similar to expected planetary components such as crustal, mantle, or core material. To understand the delivery to Earth of compounds essential for life such as water and organics, targets may include icy and volatile-rich comets as well as active and/or hydrated asteroids. These are only a few examples of the rich diversity of science objectives and associated potential targets that could be transformed into competitive sample return missions.

Collecting such a rich set of returned samples to achieve these science goals is not without challenges that must be addressed. While some samples may be "scooped" from a gravelly surface, others will require a more precise and careful collection. For instance, samples from surfaces rich in salts as observed on Ceres would provide key constraints on internal body processes as well as the initial accretion location of the body. However, experience with salt clasts in ordinary chondrite meteorites shows that the handling and analyses of such friable material is extremely challenging. As another example, samples rich in particular volatiles are missing from our meteorite collections because reentry through the Earth's atmosphere destroys such material. Volatile compounds include the water ice and organics that are essential for life on Earth, and so tracing the origin and evolution of volatile-rich materials is crucial to understanding why life arose here and the potential for life elsewhere. Successfully sampling and transporting such volatile compounds either cryogenically or non-cryogenically, in an intentionally altered state, are major spacecraft instrument design goals. Ambitious sample return missions will deliver exciting results unobtainable through other means, but will require advance investment in instrumentation and mission design.

The high science value of returned samples is enabled by the extensive and precise capabilities of Earth-based laboratories. For instance, firmly linking asteroids and meteorites requires detailed lab comparisons such as those used by Hayabusa to connect ordinary chondrites and S-type asteroids as well as by Stardust to connect comets and IDPs. There are still many well-studied meteorites whose origins are only hypothesized. For instance, identifying the parent body and origin story of iron and stony-iron achondrites is a major science objective that will require laboratory comparison of returned samples. Such knowledge would mark major progress towards the goal of understanding how growing protoplanets differentiate. Relatedly, there are a number of asteroids with rarer spectral appearances, of which it is unclear if they are represented in our meteorite collections at all, such as the A-, O-, R-, E-, M-, and non-Vestoid V-type asteroids. Hypotheses for the origin of these objects include silicate pieces of differentiated bodies, and they may be the key for understanding how the geochemistry of large protoplanets change with growth. Petrologic experiments and meteoritic comparison is only possible with returned samples and laboratory analysis.

**A Vigorous Sample Return Program is an Essential Component of Small Body Exploration**

Small body investigations began as telescopic observations and meteoritic analyses and, in the past decades, have since included spacecraft-based remote and in-situ analyses. While all of these investigations should continue to be supported by funding agencies including remote and in-situ spacecraft-based investigations, the time is now to create a New Frontiers-like program to support a set of spacecraft missions to bring back samples from a diverse set of small bodies and to increase support for Earth-based laboratories to properly handle and study returned samples. The most effective small body sample return program would have these components:
- *(1) opportunities for small body sample return spacecraft missions,*
- *(2) opportunities to develop spacecraft sampling and sample preservation technology,*
- *(3) opportunities to create or enhance existing Earth-based laboratory facilities,*
- *(4) opportunities to examine existing returned samples and/or their data products.*

The number and diversity of small body targets means that no single mission could accomplish the full set of achievable scientific objectives made possible by small body sample return missions. Ideally, sample return missions should be supported with their own mission program with its own funding line within the planetary sciences division at NASA. Alternatively, the regular calls from the Discovery and New Frontiers programs should **always** be broad enough to include small body sample return as a potential mission profile and/or commitments should be made for selection of sample return missions on a regular basis. The very strong science cases for small body sample return missions will make them highly competitive against other possible mission profiles. Given the large diversity of possible sample return science goals requiring New Frontiers-level support, it's important for that program to consider a wide variety of targets. In short, given the diversity of possible small body sample return missions, it is essential that NASA provide regular opportunities for such missions to be flown.

Since foreign space agencies, such as JAXA and ESA, will continue or start to conduct their own sample return programs, it is critical that NASA support participating scientist programs for these non-NASA led missions. Relatedly, NASA should encourage the participation of international colleagues in sample return missions as well as the Earth-based study of returned samples. Likewise, NASA should work to maintain access for US-based scientists to returned samples obtained by foreign space agencies.

NASA should continue to maintain stable, well-funded research and analysis (R&A) programs as the backbone for scientific advancement and as an enabler for future missions to small bodies. It is essential that either a new dedicated program be introduced or the PICASSO and MatISSE programs be enhanced to support the development of spacecraft sampling and onboard sample preservation technologies. As the volume of returned samples grows, NASA must increase the funding to LARS as well as preserve/increase funding to related R&A programs such as Emerging Worlds (EW) and Solar System Workings (SSW) including support for Planetary Major Equipment (PME) requests. The connective nature of returned samples is reflected in the wide number of NASA programs necessary to support their study and return their full science value. Lastly, NASA should maintain and update current facilities at Johnson Space Center and, potentially, elsewhere to store returned samples. As the variety of returned samples increases, NASA will need to develop new storage facilities, new procedures for sample handling and access. NASA should require that sample documentation, measurements, and other data products be archived appropriately, ideally at a Planetary Data System node.

**A Successful Small Body Sample Return Program Must be Accessible, Equitable, and Inclusive**

Small body sample return investigations are a human endeavor that should be accessible to anyone interested. From large spacecraft mission teams to the smaller research teams supported by R&A programs, those selected to conduct the science should represent the diversity of humanity. This can only be achieved if NASA prioritizes accessibility and inclusivity when designing competitive research programs, and if international participation is encouraged. Given historic injustices and existing inequality, NASA will need to consciously build accessibility and inclusivity into a small body sample return program. Proactive measures may include continued support of participating scientist programs to join spacecraft mission teams, maintaining transparent and fair systems to access returned samples with appropriate oversight, and preference to funding facilities that provide for guest access including training and support.

**Summary and Recommendations**

In order to tell the story of the solar system, a multi-target small body sample return program should include:
1. Opportunities for small body sample return spacecraft missions, either as a standalone program or to always include small body sample return as an admissible mission profile in the Discovery and New Frontiers programs.
2. Introduction of a competitive program for developing spacecraft sampling and sample preservation technologies or opportunities to propose these technologies under the PICASSO and MatISSE programs
3. Opportunities to create or enhance existing Earth-based laboratory facilities via increased support of LARS and the PME program
4. Regular opportunities to examine existing returned samples and/or their data products via increased support of LARS and related R&A programs such as SSW and EW
5. Proactive measures to ensure accessibility, equitableness, and inclusivity, such as participating scientist programs as well as equitable access to returned samples and user facilities